\documentclass[mathleft,fleqn,%
]{an}
\usepackage{graphicx}
\usepackage[varg]{txfonts}
\usepackage{natbib}
\bibpunct{(}{)}{;}{a}{}{,}
\begin{document}

\Pagespan{1}{}
\Yearpublication{2014}%
\Yearsubmission{2014}%
\Month{0}%
\Volume{999}%
\Issue{0}%
\DOI{asna.201400000}%

\title{Simulated X-ray Emission in Galaxy Clusters with Feedback from Active Galactic Nuclei}

\author{Rudrani Kar Chowdhury\inst{1}\fnmsep\thanks{Corresponding author:
        {rudranikarchowdhury@gmail.com}}, Soumya Roy\inst{2}, Suchetana Chatterjee\inst{1},
        Nishikanta Khandai\inst{3}, Craig L. Sarazin\inst{4}
\and  Tiziana Di Matteo\inst{5}
}
\titlerunning{X-ray emission from galaxy clusters }
\authorrunning{Kar Chowdhury et. al}
\institute{
Presidency University, Kolkata - 700073, India
\and 
Inter-University Centre for Astronomy and Astrophysics, Pune - 411007, India
\and 
National Institute of Science Education and Research, Jatni - 752050, India
\and
University of Virginia, Charlottesville, VA 22904-4325, USA
\and
Carnegie Mellon University, Pittsburgh, PA 15213, USA
}

\received{XXXX}
\accepted{XXXX}
\publonline{XXXX}


\abstract{To investigate the effect of feedback from active galactic nuclei (AGN) on their
surrounding medium, we study the diffuse X-ray emission
from galaxy groups and clusters by coupling the
Astrophysical Plasma Emission Code (APEC) with the
cosmological hydrodynamic simulation involving AGN
feedback. We construct a statistical sample of synthetic
Chandra X-ray photon maps to observationally
characterize the effect of AGN on the ambient medium. We
show that AGN are effective in displacing the hot X-ray
emitting gas from the centers of groups and clusters, and
that these signatures remain evident in observations of the
X-ray surface brightness profiles.}

\maketitle

\section{Introduction}

From different studies it is now established that active galactic nuclei (AGN) play important role in the cosmological evolution of their host galaxies and dark matter halos \citep[e.g.,][]{k&h00,DiMatteo05,Chatterjee08}. There are many evidences which indicate that the formation and evolution of the central AGN and the diffuse gas in the halo are coupled together by the activity of the supermassive black hole (SMBH), usually termed as AGN feedback in the literature \citep[e.g,][]{S&R98,Springel05b,Chatterjee15,Harrison18}. The pioneering studies involving the signature of AGN feedback was carried out by examining the effect of feedback on the distribution of diffuse X-ray emitting gas in galaxy clusters and groups \citep[e.g.,][]{Fabian06, Wise07, Mc&N07, Mc&N12}.

AGN feedback and their effects on the growth of structures have also been incorporated in cosmological simulations \citep{DiMatteo05, Sijacki07, DiMatteo08, McCarthy10, Vogelsberger14, Dubois14, Khandai15, Sijacki15, Schaye15, Steinborn15, Weinberger17, RKC19, Truong20} and they have proven to be great testbeds for comparison with observational results. Various scaling relations between the properties of the central AGN, host halo and the X-ray luminosity have been explored by several groups using different cosmological volume simulations \citep[e.g.,][]{Sijacki07, Puchwein08, Fabjan10, McCarthy10, McCarthy11, Gaspari14, LeBrun14, Barnes17, LeBrun17, Henden18, Truong18}.

In this work, we employ cosmological simulations of galaxy groups and clusters (including and excluding AGN feedback) and model the X-ray emission from them using the Astrophysical Plasma Emission Code \citep[APEC:][]{Smith01} to construct the synthetic {\it Chandra} photon maps of these systems to compare our work with observations \citep [M19 hereafter]{Chatterjee15, Mukherjee18}. Previous studies that have pioneered these techniques of mock observations were carried out by \cite{nagai07, Mendygral11, biffi12, biffi18, Wilms14, Cucchetti18}. Here, we employ the results from the high resolution fully cosmological hydrodynamical simulation with radiative gas cooling, star formation, supernova feedback and the feedback from AGN \citep{Khandai15, DiMatteo08} to generate synthetic X-ray observations. We statistically study a large sample of galaxy groups and clusters to understand the feedback effect of AGN on their surrounding hot X-ray emitting gas and compare our results with the direct observational findings of high redshift AGN.

In the next section, we discuss the basic parameters of the simulation and describe our methodology for constructing the mock photon maps. In section 3 we present our results and discuss their implications.

\section{Simulation}

We have used an extended version of the GADGET-3 simulation \citep{Springel05} for this work which is a parallel cosmological Tree Particle Mesh - Smoothed Particle Hydrodynamics code. Parameters of the simulation are consistent with the $\Lambda CDM$ cosmology. Due to limitations in spatial resolution, star-formation, BH accretion and feedback are modeled by considering the relevant physics at the sub resolution scale (see \citealt{Khandai15}, K15 hereafter, for more details). Black holes are considered as collisionless sink particles which can grow either by accreting gas that surround those or by coalescing as a result of merging. The accretion rate of gas onto the BH is estimated following the prescription of Bondi-Hoyle-Lyttleton \citep{HoyleLytt, BondiHoyle, Bondi52}. 

The bolometric luminosity of the black hole is given as 
\begin{equation}
L_{bol}=\eta\dot{M}_{BH}c^2
\end{equation}
where $c$ is the speed of light and $\eta$ is the radiative efficiency. In this simulation $\eta$ is fixed at 0.1 assuming radiatively efficient thin disk accretion \citep{ShakuraSunyaev73}. A fraction of this luminosity gets isotropically distributed to the surrounding medium (following a kernel function) as feedback energy. Rate of deposition of the energy is given by
\begin{equation}
\dot{E}_{f}=\varepsilon_{f}L_{bol}
\label{eq:Ef}
\end{equation}
where $\varepsilon_{f}$ is the feedback efficiency \citep{DiMatteo08}, the value of which is adjusted according to the normalization of the observed $M_{BH}-\sigma$ relation \citep{DiMatteo05}. 

Dark matter particles are grouped together on the run following the friends-of-friends (FOF) algorithm to form a dark matter halo. Substructures within a given halo are identified using the
\begin{scriptsize}
SUBFIND
\end{scriptsize}
algorithm \citep{Springel01}. Due to the unknown origin of SMBH and the limitation of cosmological simulation to incorporate detailed physics at every spatial scale, a seed BH of mass $\approx 10^{5}h^{-1}M_{\odot}$ is inserted in a halo if it has mass $\gtrsim 10^{10}h^{-1} M_{\odot}$ and already does not host any BH. These seed black holes are then allowed to grow via accretion or mergers. We refer the reader to \cite{DiMatteo08} for a detail discussion of formation and growth of BH in the simulation.

\begin{figure*}
\begin{center}
 \begin{tabular}{c}
     \includegraphics[width=7.5cm]{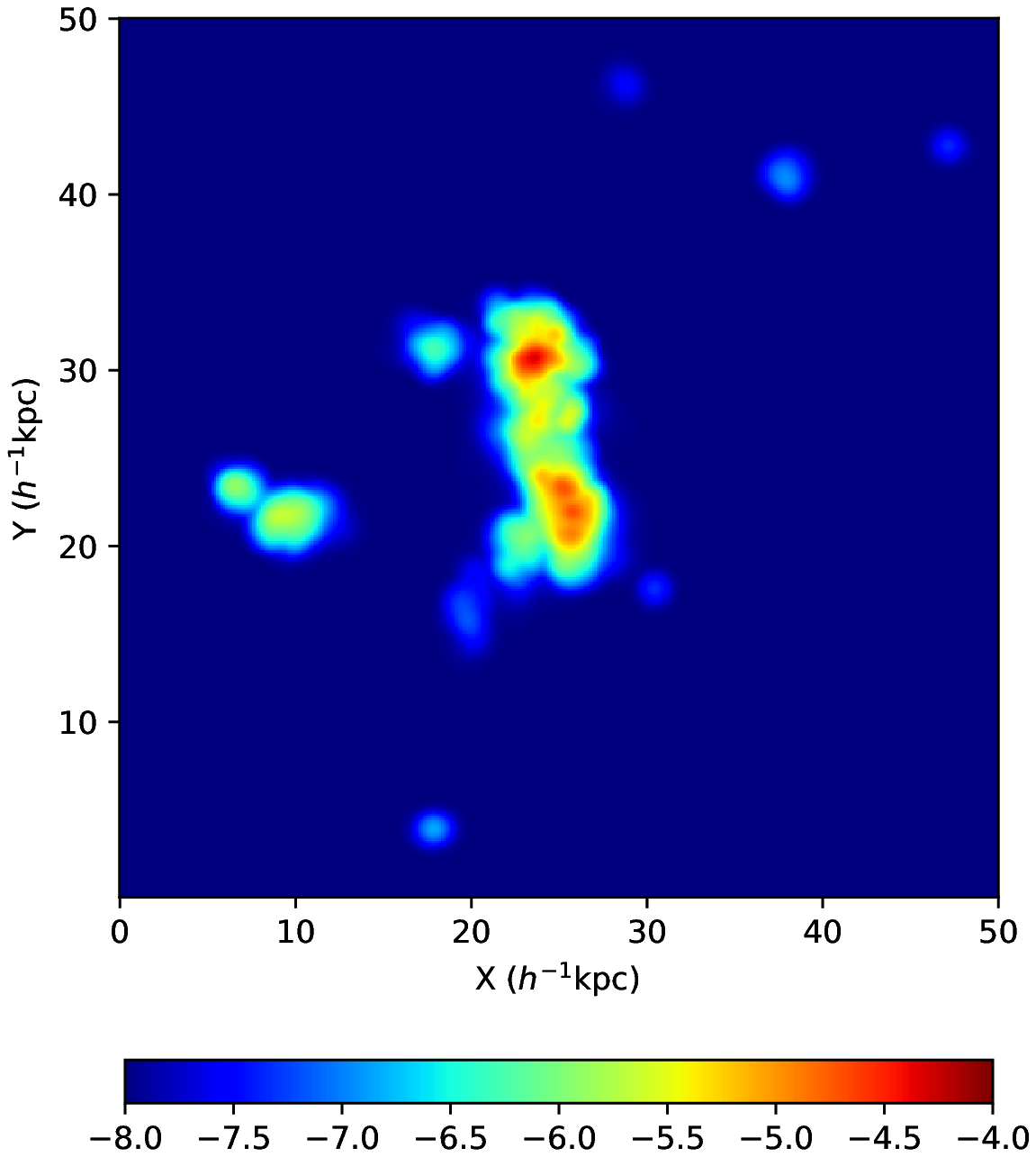}
     \includegraphics[width=7.8cm]{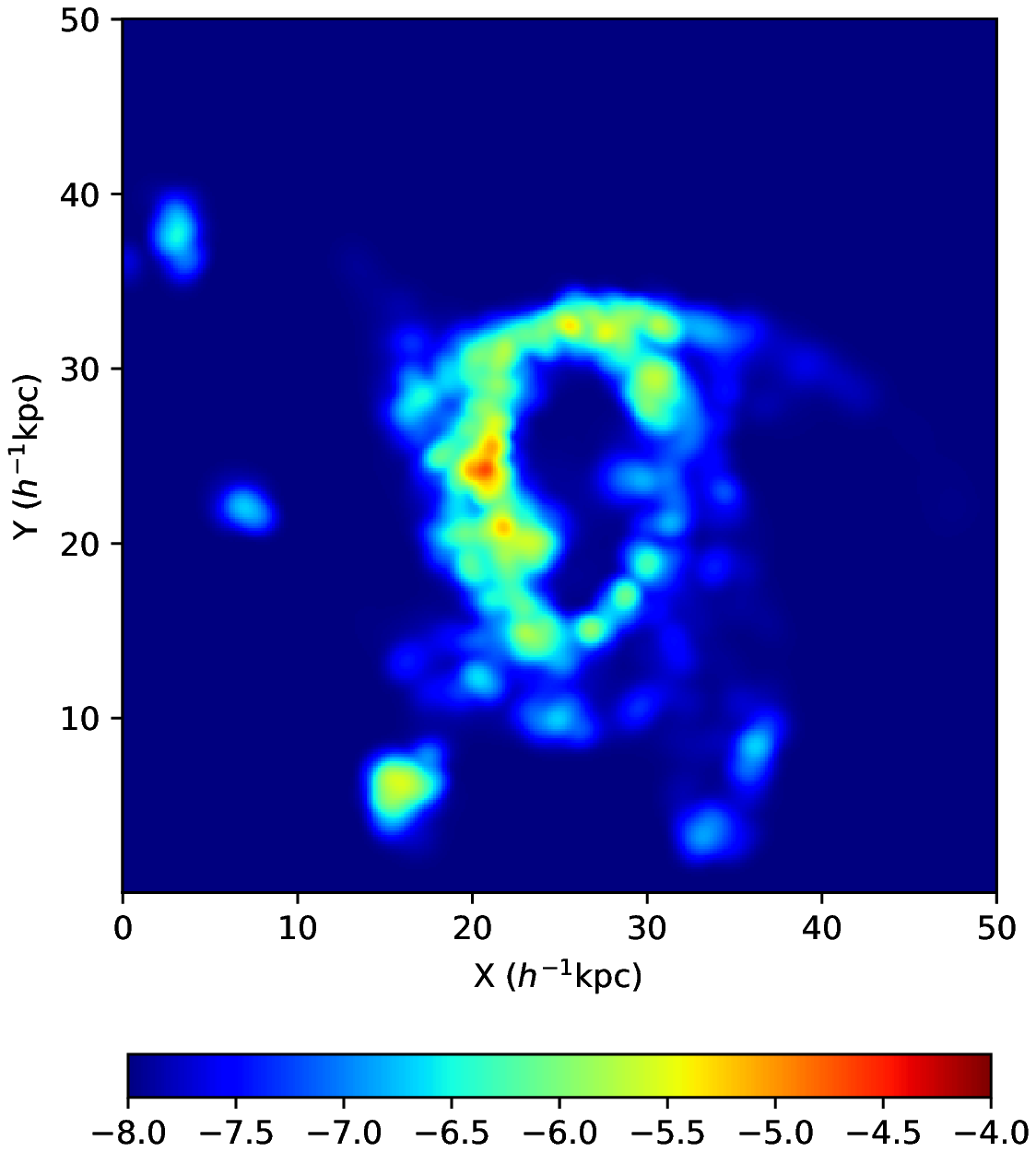}\\
     \includegraphics[width=6.5cm]{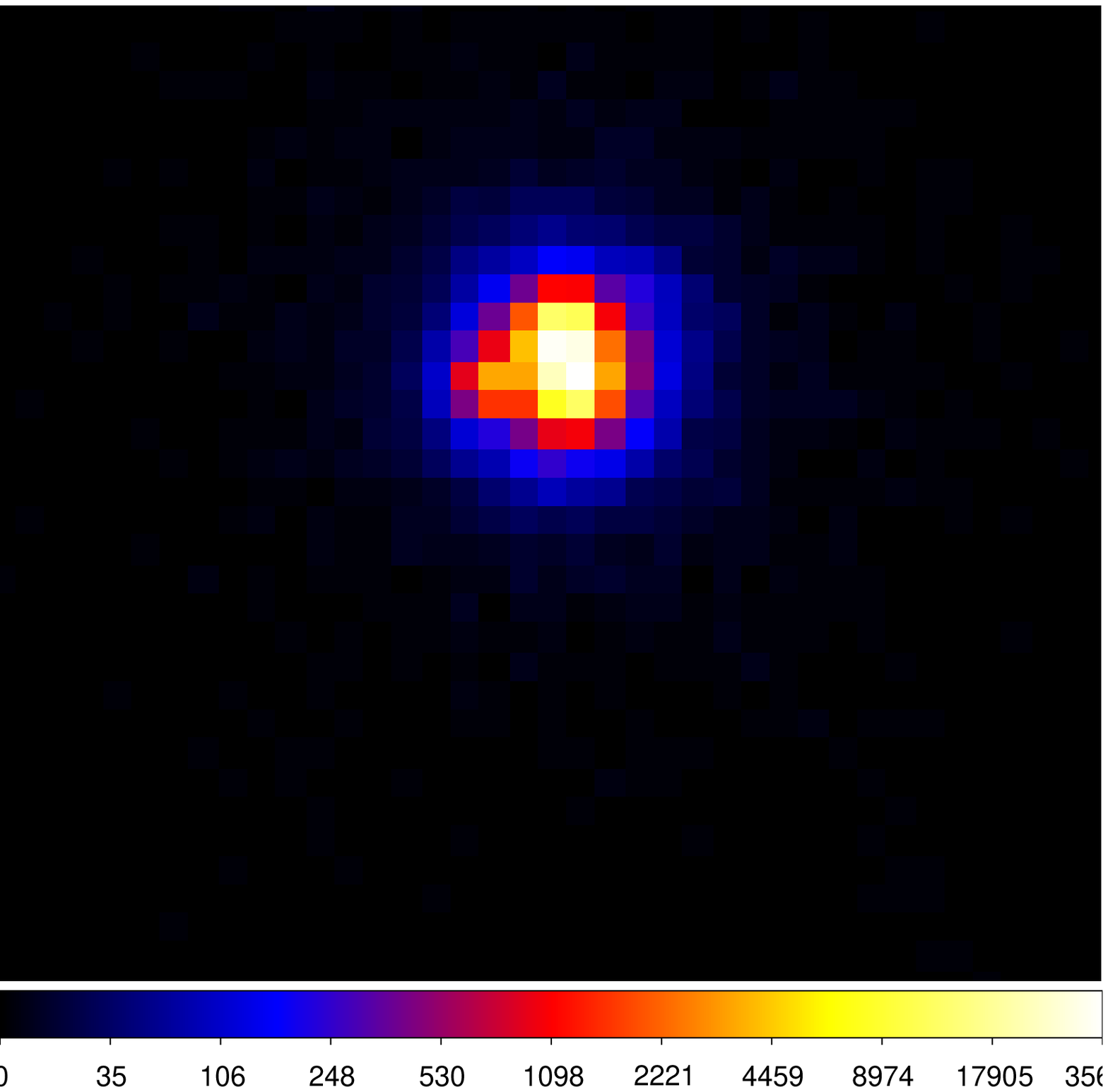}
     \includegraphics[width=6.5cm]{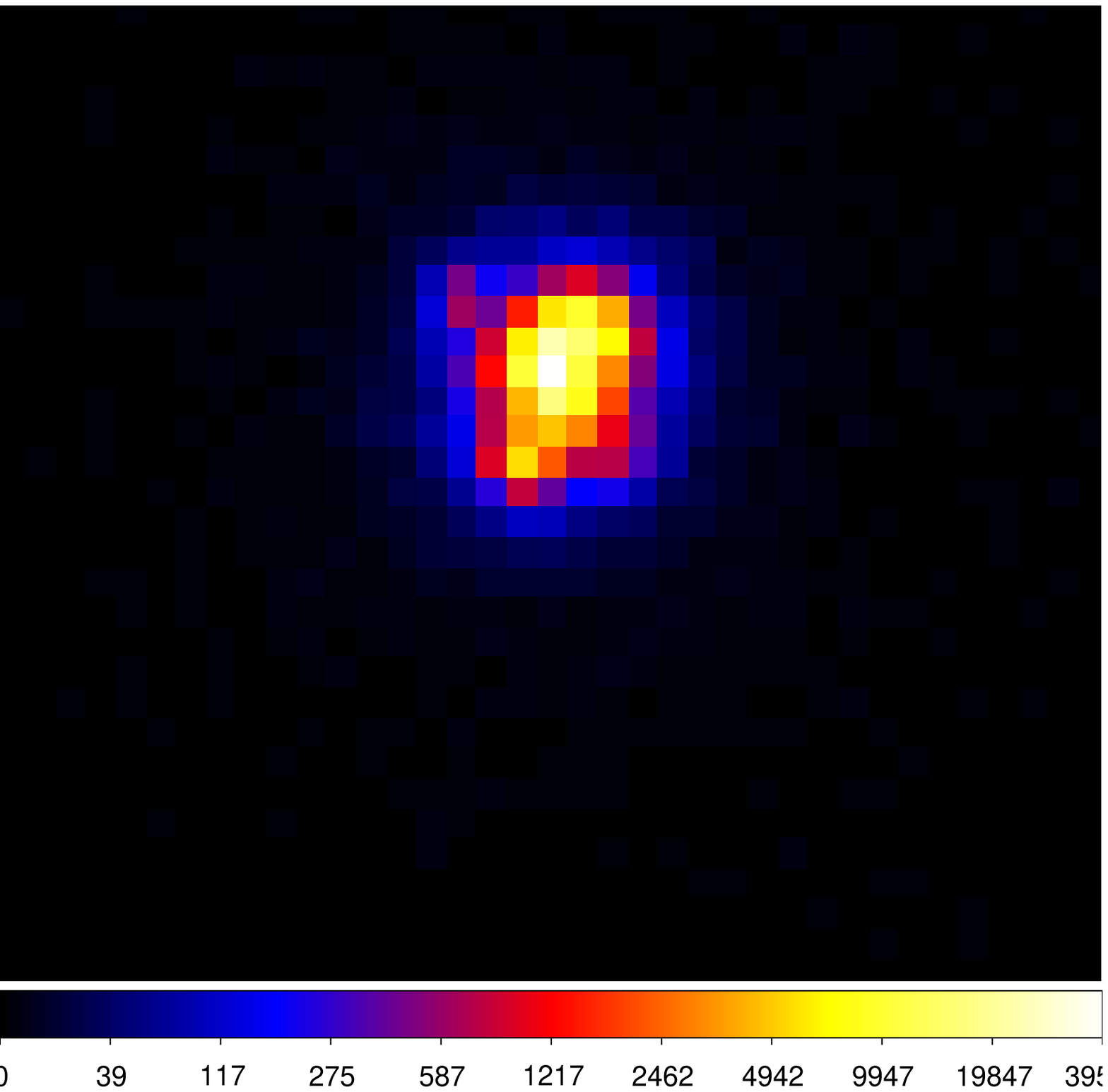}
\end{tabular}
 \caption{Images of the X-ray maps around two example AGN at K15 simulation within a region of radius 25 kpc at $z=1$. Top panel shows the theoretical X-ray map modelled by APEC. X-ray flux is color coded in the unit of $erg/cm^{2}/sec$. Bottom panel represents the corresponding mock X-ray map. Here, the color bars represent the photon number counts. All the features of the theoretical X-ray maps of the top panel is smoothed in the synthetic maps in the bottom panel due to the angular resolution of the \textit{Chandra} X-ray telescope.}
  \label{fig:map}
\end{center}
 \end{figure*}

One of the main goals of our work is to compare the X-ray environments of groups and clusters `with' and `without' AGN feedback. To make this comparison, we need two sets of simulations starting from the same initial conditions. K15 does not have a companion simulation where AGN feedback is absent. For this purpose we are using the cosmological simulation ran by \cite{DiMatteo08} [D08 hereafter] which includes both the presence and absence of AGN feedback. Both the K15 and D08 simulations have been extensively compared with observations and they have successfully reproduced the observed luminosity functions and mass functions of black holes, as well as the galaxy stellar mass function and star formation rate. We refer the reader to \cite{Khandai15}, \cite{DiMatteo08} and the references therein for the comprehensive discussion on the simulated galaxy and AGN properties and their comparison with the existing observations.

To quantify the environment of the black hole, we consider a region of radius $25 h^{-1}$ kpc around each BH and their corresponding region in without feedback simulation at z=1. This scale corresponds to an angular extent of $\sim 3''$, corresponding to the typical size of the galaxies at that particular redshift. We use the smoothed properties of the gas particles inside this region to construct the X-ray maps. Using these gas properties we now simulate the X-ray flux emitted by the hot, optically thin, colissionally ionised gas around the AGN using APEC. It is important to clarify here that we have not considered the X-rays coming from the central AGN since our goal is to study the properties of the diffuse gas. In the actual observation of the AGN feedback signal from the stacked maps, M19 also considered only the optically selected AGN in their sample and discarded the AGN that has been detected in X-rays in order to avoid contamination of the point spread function (PSF) of the central AGN.

\subsection{Construction of Synthetic X-ray Maps}

We aim to construct synthetic X-ray observations using the K15 and D08 simulation data as input. For this purpose we have used the ray tracing simulator of the onboard \textit{Chandra} X-ray Observatory (CXO) - Model of AXAF Response to X-rays \citep[MARX:][]{Davis12}. This portable simulator is designed to construct the {\it Chandra} event files. It considers all the phases that a photon undergoes from incident on the mirror of the telescope to its detection by the detector, including detailed models of the High Resolution Mirror Assembly (HRMA) and the Advanced CCD Imaging Spectrometer (ACIS) of the spacecraft. We have used the parameter files dated 2009 July 1 for our synthetic observation.

We use the Chandra Interactive Analysis of Observations (CIAO) software \citep{Fruscione06} for the analysis of mock event files. CIAO module \textit{pycrates} and \textit{dmhedit} are used to convert the format of the input files into MARX specific format and assign angular scales to the maps using pseudo World Co-ordinate System (WCS). We now simulate the photons using MARX for 200 ksec exposure time to match our sample with the observational data of M19. We repeat the same procedure for all the BHs present in our simulation at z=1 and then co-add them to get the stacked X-ray maps for all the AGN with the help of another CIAO module \textit{dmmerge}. One of the main motivation for stacking is to statistically study the effect of AGN feedback on its surrounding hot X-ray emitting gas. While AGN feedback in the individual systems might not be detected due to limited spatial resolution of the telescope involved, we can expect to detect the signature of feedback in the stacked X-ray maps. We create radial profiles of the X-ray flux from stacked X-ray maps in presence and absence of AGN feedback to explore the effect of AGN feedback on cluster X-ray emission. This has been done with the help of SAOImage DS9, which is an astronomical imaging and data visualization software.

\section{Results and Discussions}

In this work we focus on the diffuse X-ray emission in galaxy groups and clusters in the presence and absence of AGN feedback to study the signature of feedback on the surrounding medium of the AGN using a full cosmological volume simulation. Fig. \ref{fig:map} shows the X-ray emission around two example AGN in K15 simulation within a region of radius $25 h^{-1}$ kpc at $z=1$. We model the X-ray emission from the simulated galaxy clusters using the plasma emission code APEC. This is shown in the top panel. Here the X-ray flux is color coded in the unit of $erg/cm^2/sec$. Bottom panel shows the corresponding synthetic X-ray maps of \textit{Chandra} telescope. Here, the colorbars represent the photon number counts. Black holes are located at the centre of each image. As noted before, we do not compute the X-ray emission coming from the central black hole in this analysis. We later compare the X-ray emission with the case where AGN feedback is absent and show that AGN play significant role in evacuating gas from the central regions of the galaxy groups and clusters. However, the synthetic X-ray maps appear to be extremely smooth and no features of the theoretical maps can be seen there. We understand this is due to the PSF of the \textit{Chandra} telescope. We later discuss the effect of this smoothing in our analysis.

\begin{figure*}
\begin{center}
\includegraphics[width=14cm]{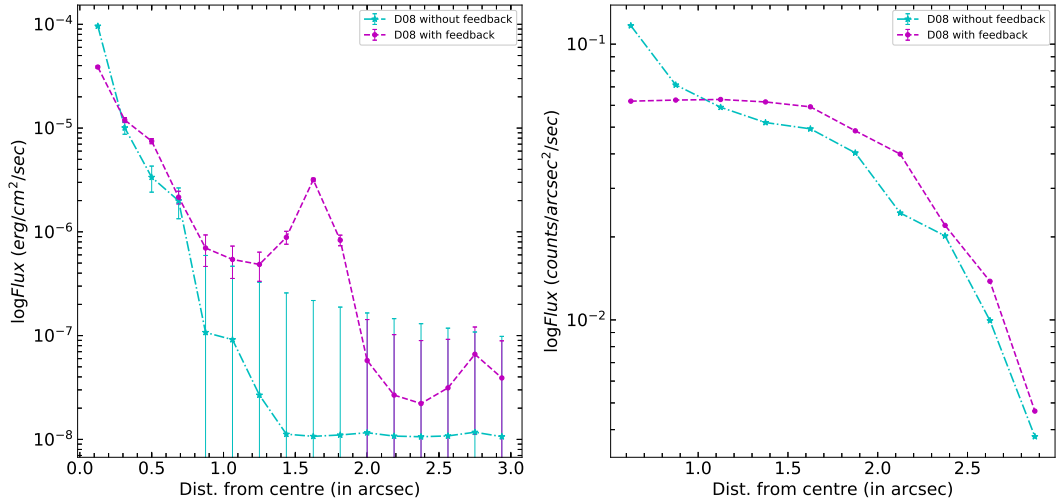}
\caption{Stacked radial profile of the X-ray flux of the diffuse gas in the presence and absence of AGN feedback in D08 simulation at $z=1$. Left panel: Stacked radial profile of theoretical X-ray flux of diffuse gas around AGN of $L_{bol}>10^{43}$ erg/sec and their without feedback counterpart. Right panel: Stacked radial profile of mock X-ray flux around the same sample of AGN. Cyan stars and magenta dots represent the without and with feedback scenario respectively in both the panels. Central deficit of X-ray flux in the presence of AGN feedback is visible in both theoretical and synthetic observation.}
\label{fig:profile}
\end{center}
\end{figure*}

To statistically study the signature of AGN feedback activity on the diffuse X-ray emitting gas, we consider D08 `with' and `without' feedback simulation at $z=1$. We select the AGN having $L_{bol} > 10^{43}$ erg/sec and their counterpart in the without feedback simulation to qualitatively match our sample with M19 dataset. Total number of such system is 161. We construct X-ray maps for all the systems and stack them. Finally, we examine the radial profile of the stacked X-ray maps in the presence and absence of AGN feedback. Fig. \ref{fig:profile} shows stacked radial profiles of the X-ray flux around the black holes in the presence and absence of AGN feedback. Left panel is purely theoretical profile while the right panel depicts the radial profile of the stacked photon map simulated using MARX. Error bars on the data points are the standard errors at each annulus. Cyan stars and magenta dots in both the panels represent the scenario when feedback is absent and present in the D08 simulation, respectively. Here, we note that, the best spatial resolution of \textit{Chandra} X-ray telescope is $0.5''$. Therefore, any difference of the X-ray flux inside this scale is not resolved in the stacked synthetic profile in the right panel.

The central excess of the X-ray flux in the absence of feedback activity is clearly noticeable from both the theoretical and synthetic profiles. This agrees with the claims made in other previous studies that AGN feedback is displacing the surrounding gas outwards. \citet{Gaspari11} studied the properties of the outflow from AGN at galaxy group scales using 3D adaptive mesh refinement (AMR) simulation. Their study shows low surface brightness of X-ray flux at the central region, similar to our finding of the X-ray underluminous region at the centre. \citet{Pellegrini12} did a detailed hydrodynamic simulation of AGN feedback to study the association of different properties of ISM of elliptical galaxies with the nuclear activity. They found that the X-ray surface brightness profile decreases at the central part of galaxies in presence of the AGN outburst which is also consistent with our results. This is also observed from the actual observational results. M19 performed a stacking analysis of the X-ray emission from $(z \approx 0.6) $ normal and active galaxies to test for AGN feedback activity in the ICM/interstellar medium (ISM). They found that the central region of the stacked X-ray map is more luminous in X-rays for the normal galaxies when compared with AGN. Our results are in agreement with their findings too, suggesting that the feedback from AGN is indeed expelling the X-ray emitting gas away from the central region.

We can also calculate the integrated difference between the X-ray surface brightness profiles in the presence and absence of AGN feedback following the approach mentioned in \citet{Chatterjee15}. The excess flux can be calculated as 
\begin{equation}
\bigtriangleup F(r) = F_{without}(r) - F_{with}(r)
\label{eq:Excess},
\end{equation}
where $F_{without}(r)$ and $F_{with}(r)$ represent X-ray flux in the absence and presence of AGN feedback at a distance $r$ from the centre.
Furthermore, we can also calculate the fraction of the bolometric energy contributing to the feedback of AGN by measuring the additional energy in the absence of AGN feedback from the integrated difference in Eq \ref{eq:Excess}. It is possible to estimate the approximate energy difference in the presence and absence of AGN feedback following the scheme mentioned in \S 4 of \citet{Chatterjee15}. An estimate of the excess energy can be calculated as $\bigtriangleup E = 4\pi d_{L}^{2}(\int 2\pi r \bigtriangleup F(r) dr)$ , where $d_{L}$ is the luminosity distance. We plan to do this analysis in future using our theoretical and synthetic X-ray flux in the presence and absence of AGN feedback and compare it with the observed value of M19 in order to shed more light on the instrumental effect on any observation.

Our current study also allows us to characterize the instrumental effect of {\it Chandra} in particular, and we see that the X-ray features are flattened in the right panel when compared to the left panel of Fig. \ref{fig:profile}. We further plan to calculate the feedback fraction, which is the ratio of the feedback energy to the total radiated energy of the AGN and compare it with the findings of M19. We propose to check this method with multiple simulations and observations, for stronger validation and characterization of the instrumental response function in interpreting the observational results. We further propose to undertake this characterization and synthetic observation scheme with future missions such as \textit{Athena} (Kar Chowdhury et al. in preparation).

Finally, it is important to mention the limitations of the simulation used in this work. First of all, the star formation and the supernova feedback as well as the formation and the growth of the black holes are based on the subgrid model in both the D08 and K15 simulation due to the limited computational resources available. But we mention that as long as the large scale effect of the AGN feedback matches with the observation, a subgrid model can be a fair representation of the detailed physics at the small scale. Also, the spherical Bondi model is adapted in the simulation, instead of the rigorous relativistic accretion model, to constrain the accretion onto the black hole. Besides, only radiative mode of AGN feedback is assumed in the simulation. We refer the reader to \citet{DiMatteo08} for a thorough discussion on this. However, in a future work we plan to consider both the kinetic and radiative mode of feedback using the Illustris-TNG simulation \citep{Weinberger17}.

In our work, we have used one of the highest resolution cosmological hydrodynamic simulation (MassiveBlackII) involving detailed modeling of radiative gas cooling, star formation, supernova feedback and the feedback from AGN to generate a large statistical sample of synthetic X-ray observations of galaxy groups and clusters. We show the effect of feedback on X-ray images and profiles of groups and clusters with active black holes at their centers and validate the method of X-ray stacking in detecting feedback from systems that can not be resolved with current X-ray telescopes. AGN feedback has been an important component in understanding galaxy evolution as well as cluster formation. Through our numerical and statistical work we demonstrate the new observational directions in which we can study this interesting phenomenon through current and upcoming missions.

\acknowledgements
  
RKC thanks Chandra X-ray Observatory helpdesk system, Dr. Nicholas P. Lee and Dr. Antonella Fruscione for helping us to work with the CIAO software package. This research has made use of the software provided by the Chandra X-ray Center (CXC) in the application packages CIAO. SC acknowledges financial support from the Department of Science and Technology (DST) through the SERB Early Career Research grant and Presidency University through the Faculty Research and Professional Development Funds. SC and NK are grateful to the Inter University Center for Astronomy and Astrophysics (IUCAA) for providing infra-structural and financial support along with local hospitality through the IUCAA-associateship program. NK is supported by the Ramanujan Fellowship awarded by the DST, Government of India. The simulations were run on the Cray XT5 supercomputer Kraken at the National Institute for Computational Sciences. The analysis was partially done on the xanadu cluster funded by the Ramanujan Fellowship at NISER. NK and TDM acknowledge support from National Science Foundation (NSF) PetaApps program, OCI-0749212 and by NSF AST-1009781.  

\bibliographystyle{an}
\bibliography{ref}

\begin{thebibliography}{50}
\expandafter\ifx\csname natexlab\endcsname\relax\def\natexlab#1{#1}\fi

\bibitem[{{Barnes} {et~al.}(2017){Barnes}, {Kay}, {Bah{\'e}}, {Dalla Vecchia},
  {McCarthy}, {Schaye}, {Bower}, {Jenkins}, {Thomas}, {Schaller}, {Crain},
  {Theuns}, \& {White}}]{Barnes17}
{Barnes}, D.~J., {Kay}, S.~T., {Bah{\'e}}, Y.~M., {et~al.} 2017, \mnras, 471,
  1088

\bibitem[{{Biffi} {et~al.}(2012){Biffi}, {Dolag}, {B{\"o}hringer}, \&
  {Lemson}}]{biffi12}
{Biffi}, V., {Dolag}, K., {B{\"o}hringer}, H., \& {Lemson}, G. 2012, \mnras,
  420, 3545

\bibitem[{{Biffi} {et~al.}(2018){Biffi}, {Dolag}, \& {Merloni}}]{biffi18}
{Biffi}, V., {Dolag}, K., \& {Merloni}, A. 2018, \mnras, 481, 2213

\bibitem[{{Bondi}(1952)}]{Bondi52}
{Bondi}, H. 1952, \mnras, 112, 195

\bibitem[{{Bondi} \& {Hoyle}(1944)}]{BondiHoyle}
{Bondi}, H. \& {Hoyle}, F. 1944, \mnras, 104, 273

\bibitem[{{Chatterjee} {et~al.}(2008){Chatterjee}, {Di Matteo}, {Kosowsky}, \&
  {Pelupessy}}]{Chatterjee08}
{Chatterjee}, S., {Di Matteo}, T., {Kosowsky}, A., \& {Pelupessy}, I. 2008,
  \mnras, 390, 535

\bibitem[{{Chatterjee} {et~al.}(2015){Chatterjee}, {Newman}, {Jeltema},
  {Myers}, {Aird}, {Coil}, {Cooper}, {Finoguenov}, {Laird}, {Montero-Dorta},
  {Nandra}, {Willmer}, \& {Yan}}]{Chatterjee15}
{Chatterjee}, S., {Newman}, J.~A., {Jeltema}, T., {et~al.} 2015, \pasp, 127,
  716

\bibitem[{{Cucchetti} {et~al.}(2018){Cucchetti}, {Pointecouteau}, {Peille},
  {Clerc}, {Rasia}, {Biffi}, {Borgani}, {Tornatore}, {Dolag}, {Roncarelli},
  {Gaspari}, {Ettori}, {Bulbul}, {Dauser}, {Wilms}, {Pajot}, \&
  {Barret}}]{Cucchetti18}
{Cucchetti}, E., {Pointecouteau}, E., {Peille}, P., {et~al.} 2018, \aap, 620,
  A173

\bibitem[{{Davis} {et~al.}(2012){Davis}, {Bautz}, {Dewey}, {Heilmann}, {Houck},
  {Huenemoerder}, {Marshall}, {Nowak}, {Schattenburg}, {Schulz}, \&
  {Smith}}]{Davis12}
{Davis}, J.~E., {Bautz}, M.~W., {Dewey}, D., {et~al.} 2012, in \procspie, Vol.
  8443, Space Telescopes and Instrumentation 2012: Ultraviolet to Gamma Ray,
  84431A

\bibitem[{{Di Matteo} {et~al.}(2008){Di Matteo}, {Colberg}, {Springel},
  {Hernquist}, \& {Sijacki}}]{DiMatteo08}
{Di Matteo}, T., {Colberg}, J., {Springel}, V., {Hernquist}, L., \& {Sijacki},
  D. 2008, \apj, 676, 33

\bibitem[{{Di Matteo} {et~al.}(2005){Di Matteo}, {Springel}, \&
  {Hernquist}}]{DiMatteo05}
{Di Matteo}, T., {Springel}, V., \& {Hernquist}, L. 2005, \nat, 433, 604

\bibitem[{{Dubois} {et~al.}(2014){Dubois}, {Pichon}, {Welker}, {Le Borgne},
  {Devriendt}, {Laigle}, {Codis}, {Pogosyan}, {Arnouts}, {Benabed}, {Bertin},
  {Blaizot}, {Bouchet}, {Cardoso}, {Colombi}, {de Lapparent}, {Desjacques},
  {Gavazzi}, {Kassin}, {Kimm}, {McCracken}, {Milliard}, {Peirani}, {Prunet},
  {Rouberol}, {Silk}, {Slyz}, {Sousbie}, {Teyssier}, {Tresse}, {Treyer},
  {Vibert}, \& {Volonteri}}]{Dubois14}
{Dubois}, Y., {Pichon}, C., {Welker}, C., {et~al.} 2014, Monthly Notices of the
  Royal Astronomical Society, 444, 1453

\bibitem[{{Fabian} {et~al.}(2006){Fabian}, {Sanders}, {Taylor}, {Allen},
  {Crawford}, {Johnstone}, \& {Iwasawa}}]{Fabian06}
{Fabian}, A.~C., {Sanders}, J.~S., {Taylor}, G.~B., {et~al.} 2006, \mnras, 366,
  417

\bibitem[{{Fabjan} {et~al.}(2010){Fabjan}, {Borgani}, {Tornatore}, {Saro},
  {Murante}, \& {Dolag}}]{Fabjan10}
{Fabjan}, D., {Borgani}, S., {Tornatore}, L., {et~al.} 2010, Monthly Notices of
  the Royal Astronomical Society, 401, 1670

\bibitem[{{Fruscione} {et~al.}(2006){Fruscione}, {McDowell}, {Allen},
  {Brickhouse}, {Burke}, {Davis}, {Durham}, {Elvis}, {Galle}, {Harris},
  {Huenemoerder}, {Houck}, {Ishibashi}, {Karovska}, {Nicastro}, {Noble},
  {Nowak}, {Primini}, {Siemiginowska}, {Smith}, \& {Wise}}]{Fruscione06}
{Fruscione}, A., {McDowell}, J.~C., {Allen}, G.~E., {et~al.} 2006, in
  \procspie, Vol. 6270, Society of Photo-Optical Instrumentation Engineers
  (SPIE) Conference Series, 62701V

\bibitem[{{Gaspari} {et~al.}(2011){Gaspari}, {Brighenti}, {D'Ercole}, \&
  {Melioli}}]{Gaspari11}
{Gaspari}, M., {Brighenti}, F., {D'Ercole}, A., \& {Melioli}, C. 2011, Monthly
  Notices of the Royal Astronomical Society, 415, 1549

\bibitem[{{Gaspari} {et~al.}(2014){Gaspari}, {Brighenti}, {Temi}, \&
  {Ettori}}]{Gaspari14}
{Gaspari}, M., {Brighenti}, F., {Temi}, P., \& {Ettori}, S. 2014, The
  Astrophysical Journal, 783, L10

\bibitem[{{Harrison} {et~al.}(2018){Harrison}, {Costa}, {Tadhunter},
  {Fl{\"u}tsch}, {Kakkad}, {Perna}, \& {Vietri}}]{Harrison18}
{Harrison}, C.~M., {Costa}, T., {Tadhunter}, C.~N., {et~al.} 2018, Nature
  Astronomy, 2, 198

\bibitem[{{Henden} {et~al.}(2018){Henden}, {Puchwein}, {Shen}, \&
  {Sijacki}}]{Henden18}
{Henden}, N.~A., {Puchwein}, E., {Shen}, S., \& {Sijacki}, D. 2018, Monthly
  Notices of the Royal Astronomical Society, 479, 5385

\bibitem[{{Hoyle} \& {Lyttleton}(1939)}]{HoyleLytt}
{Hoyle}, F. \& {Lyttleton}, R.~A. 1939, Proceedings of the Cambridge
  Philosophical Society, 35, 405

\bibitem[{Kar~Chowdhury {et~al.}(2020)Kar~Chowdhury, Chatterjee, Lonappan,
  Khandai, \& Di~Matteo}]{RKC19}
Kar~Chowdhury, R., Chatterjee, S., Lonappan, A.~I., Khandai, N., \& Di~Matteo,
  T. 2020, The Astrophysical Journal, 889, 60

\bibitem[{{Kauffmann} \& {Haehnelt}(2000)}]{k&h00}
{Kauffmann}, G. \& {Haehnelt}, M. 2000, \mnras, 311, 576

\bibitem[{{Khandai} {et~al.}(2015){Khandai}, {Di Matteo}, {Croft}, {Wilkins},
  {Feng}, {Tucker}, {DeGraf}, \& {Liu}}]{Khandai15}
{Khandai}, N., {Di Matteo}, T., {Croft}, R., {et~al.} 2015, \mnras, 450, 1349

\bibitem[{{Le Brun} {et~al.}(2014){Le Brun}, {McCarthy}, {Schaye}, \&
  {Ponman}}]{LeBrun14}
{Le Brun}, A. M.~C., {McCarthy}, I.~G., {Schaye}, J., \& {Ponman}, T.~J. 2014,
  \mnras, 441, 1270

\bibitem[{{Le Brun} {et~al.}(2017){Le Brun}, {McCarthy}, {Schaye}, \&
  {Ponman}}]{LeBrun17}
{Le Brun}, A. M.~C., {McCarthy}, I.~G., {Schaye}, J., \& {Ponman}, T.~J. 2017,
  \mnras, 466, 4442

\bibitem[{{McCarthy} {et~al.}(2011){McCarthy}, {Schaye}, {Bower}, {Ponman},
  {Booth}, {Dalla Vecchia}, \& {Springel}}]{McCarthy11}
{McCarthy}, I.~G., {Schaye}, J., {Bower}, R.~G., {et~al.} 2011, Monthly Notices
  of the Royal Astronomical Society, 412, 1965

\bibitem[{{McCarthy} {et~al.}(2010){McCarthy}, {Schaye}, {Ponman}, {Bower},
  {Booth}, {Dalla Vecchia}, {Crain}, {Springel}, {Theuns}, \&
  {Wiersma}}]{McCarthy10}
{McCarthy}, I.~G., {Schaye}, J., {Ponman}, T.~J., {et~al.} 2010, \mnras, 406,
  822

\bibitem[{{McNamara} \& {Nulsen}(2007)}]{Mc&N07}
{McNamara}, B.~R. \& {Nulsen}, P.~E.~J. 2007, \araa, 45, 117

\bibitem[{{McNamara} \& {Nulsen}(2012)}]{Mc&N12}
{McNamara}, B.~R. \& {Nulsen}, P.~E.~J. 2012, New Journal of Physics, 14,
  055023

\bibitem[{{Mendygral} {et~al.}(2011){Mendygral}, {O'Neill}, \&
  {Jones}}]{Mendygral11}
{Mendygral}, P.~J., {O'Neill}, S.~M., \& {Jones}, T.~W. 2011, \apj, 730, 100

\bibitem[{{Mukherjee} {et~al.}(2019){Mukherjee}, {Bhattacharjee}, {Chatterjee},
  {Newman}, \& {Yan}}]{Mukherjee18}
{Mukherjee}, S., {Bhattacharjee}, A., {Chatterjee}, S., {Newman}, J.~A., \&
  {Yan}, R. 2019, \apj, 872, 35

\bibitem[{{Nagai} {et~al.}(2007){Nagai}, {Vikhlinin}, \& {Kravtsov}}]{nagai07}
{Nagai}, D., {Vikhlinin}, A., \& {Kravtsov}, A.~V. 2007, \apj, 655, 98

\bibitem[{{Pellegrini} {et~al.}(2012){Pellegrini}, {Ciotti}, \&
  {Ostriker}}]{Pellegrini12}
{Pellegrini}, S., {Ciotti}, L., \& {Ostriker}, J.~P. 2012, \apj, 744, 21

\bibitem[{{Puchwein} {et~al.}(2008){Puchwein}, {Sijacki}, \&
  {Springel}}]{Puchwein08}
{Puchwein}, E., {Sijacki}, D., \& {Springel}, V. 2008, The Astrophysical
  Journal, 687, L53

\bibitem[{{Schaye} {et~al.}(2015){Schaye}, {Crain}, {Bower}, {Furlong},
  {Schaller}, {Theuns}, {Dalla Vecchia}, {Frenk}, {McCarthy}, {Helly},
  {Jenkins}, {Rosas-Guevara}, {White}, {Baes}, {Booth}, {Camps}, {Navarro},
  {Qu}, {Rahmati}, {Sawala}, {Thomas}, \& {Trayford}}]{Schaye15}
{Schaye}, J., {Crain}, R.~A., {Bower}, R.~G., {et~al.} 2015, \mnras, 446, 521

\bibitem[{{Shakura} \& {Sunyaev}(1973)}]{ShakuraSunyaev73}
{Shakura}, N.~I. \& {Sunyaev}, R.~A. 1973, \aap, 24, 337

\bibitem[{{Sijacki} {et~al.}(2007){Sijacki}, {Springel}, {Di Matteo}, \&
  {Hernquist}}]{Sijacki07}
{Sijacki}, D., {Springel}, V., {Di Matteo}, T., \& {Hernquist}, L. 2007,
  \mnras, 380, 877

\bibitem[{{Sijacki} {et~al.}(2015){Sijacki}, {Vogelsberger}, {Genel},
  {Springel}, {Torrey}, {Snyder}, {Nelson}, \& {Hernquist}}]{Sijacki15}
{Sijacki}, D., {Vogelsberger}, M., {Genel}, S., {et~al.} 2015, \mnras, 452, 575

\bibitem[{{Silk} \& {Rees}(1998)}]{S&R98}
{Silk}, J. \& {Rees}, M.~J. 1998, \aap, 331, L1

\bibitem[{{Smith} {et~al.}(2001){Smith}, {Brickhouse}, {Liedahl}, \&
  {Raymond}}]{Smith01}
{Smith}, R.~K., {Brickhouse}, N.~S., {Liedahl}, D.~A., \& {Raymond}, J.~C.
  2001, \apjl, 556, L91

\bibitem[{{Springel}(2005)}]{Springel05}
{Springel}, V. 2005, \mnras, 364, 1105

\bibitem[{{Springel} {et~al.}(2005){Springel}, {Di Matteo}, \&
  {Hernquist}}]{Springel05b}
{Springel}, V., {Di Matteo}, T., \& {Hernquist}, L. 2005, \mnras, 361, 776

\bibitem[{{Springel} {et~al.}(2001){Springel}, {White}, {Tormen}, \&
  {Kauffmann}}]{Springel01}
{Springel}, V., {White}, S.~D.~M., {Tormen}, G., \& {Kauffmann}, G. 2001,
  \mnras, 328, 726

\bibitem[{{Steinborn} {et~al.}(2015){Steinborn}, {Dolag}, {Hirschmann},
  {Prieto}, \& {Remus}}]{Steinborn15}
{Steinborn}, L.~K., {Dolag}, K., {Hirschmann}, M., {Prieto}, M.~A., \& {Remus},
  R.-S. 2015, \mnras, 448, 1504

\bibitem[{{Truong} {et~al.}(2020){Truong}, {Pillepich}, {Werner}, {Nelson},
  {Lakhchaura}, {Weinberger}, {Springel}, {Vogelsberger}, \&
  {Hernquist}}]{Truong20}
{Truong}, N., {Pillepich}, A., {Werner}, N., {et~al.} 2020, \mnras, 494, 549

\bibitem[{{Truong} {et~al.}(2018){Truong}, {Rasia}, {Mazzotta}, {Planelles},
  {Biffi}, {Fabjan}, {Beck}, {Borgani}, {Dolag}, {Gaspari}, {Granato},
  {Murante}, {Ragone-Figueroa}, \& {Steinborn}}]{Truong18}
{Truong}, N., {Rasia}, E., {Mazzotta}, P., {et~al.} 2018, \mnras, 474, 4089

\bibitem[{{Vogelsberger} {et~al.}(2014){Vogelsberger}, {Genel}, {Springel},
  {Torrey}, {Sijacki}, {Xu}, {Snyder}, {Bird}, {Nelson}, \&
  {Hernquist}}]{Vogelsberger14}
{Vogelsberger}, M., {Genel}, S., {Springel}, V., {et~al.} 2014, \nat, 509, 177

\bibitem[{{Weinberger} {et~al.}(2017){Weinberger}, {Springel}, {Hernquist},
  {Pillepich}, {Marinacci}, {Pakmor}, {Nelson}, {Genel}, {Vogelsberger},
  {Naiman}, \& {Torrey}}]{Weinberger17}
{Weinberger}, R., {Springel}, V., {Hernquist}, L., {et~al.} 2017, \mnras, 465,
  3291

\bibitem[{{Wilms} {et~al.}(2014){Wilms}, {Brand}, {Barret}, {Beuchert}, {den
  Herder}, {Kreykenbohm}, {Lotti}, {Meidinger}, {Nand ra}, {Peille}, {Piro},
  {Rau}, {Schmid}, {Smith}, {Tenzer}, {Wille}, \& {Willingale}}]{Wilms14}
{Wilms}, J., {Brand}, T., {Barret}, D., {et~al.} 2014, Society of Photo-Optical
  Instrumentation Engineers (SPIE) Conference Series, Vol. 9144, {ATHENA
  end-to-end simulations}, 91445X

\bibitem[{{Wise} {et~al.}(2007){Wise}, {McNamara}, {Nulsen}, {Houck}, \&
  {David}}]{Wise07}
{Wise}, M.~W., {McNamara}, B.~R., {Nulsen}, P.~E.~J., {Houck}, J.~C., \&
  {David}, L.~P. 2007, \apj, 659, 1153

\end{thebibliography}

\end{document}